\documentclass[10pt]{amsart}

\usepackage{amsthm,amsmath,amssymb}
\newcommand{\F}{\mathbb{F}}
\newcommand{\E}{\mathbb{E}}
\newcommand{\Z}{\mathbb{Z}}

\newcommand{\ga}{\alpha}
\newcommand{\gb}{\beta}

\newcommand{\gd}{\delta}

\renewcommand{\phi}{\varphi}

\newcommand{\set}[1]{\{#1\}}

\newcommand{\fl}[1]{\lfloor{#1}\rfloor}
\newcommand{\ydeg}{\text{$y$-$\deg$}}

\DeclareMathOperator{\ev}{ev}
\DeclareMathOperator{\wt}{wt}
\DeclareMathOperator{\RS}{RS}
\DeclareMathOperator{\GRS}{GRS}
\DeclareMathOperator{\BCH}{BCH}
\DeclareMathOperator{\ALT}{C}
\DeclareMathOperator{\LT}{lt}
\DeclareMathOperator{\LC}{lc}

\newtheorem{thm}{Theorem}
\newtheorem{prop}[thm]{Proposition}

\theoremstyle{definition}

\newtheorem*{case1}{Case: $\deg(D)\ge\deg(B)$\normalfont}
\newtheorem*{case2}{Case: $\deg(D)<\deg(B)$\normalfont}
\newtheorem*{algD}{Decoding Algorithm D}
\newtheorem*{algE}{Euclidean Decoding Algorithm E}

\begin{document}

\title{Interpolation-based Decoding of Alternant Codes}

\author{Kwankyu~Lee}
\address{Korea Institute of Advanced Study, Seoul, Korea}
\email{kwankyu@kias.re.kr}


\begin{abstract}
We formulate the classical decoding algorithm of alternant codes afresh based on interpolation as in Sudan's list decoding of Reed-Solomon codes, and thus get rid of the key equation and the linear recurring sequences in the theory. The result is a streamlined exposition of the decoding algorithm using a bit of the theory of Gr\"obner bases of modules.
\end{abstract}

\maketitle

\section{Introduction}

The family of alternant codes embraces BCH codes and Reed-Solomon codes, which are important in the practice of error control coding. The popular decoding algorithm of BCH codes using the Berlekamp-Massey algorithm or the Euclidean algorithm in fact decodes any alternant code for errors of weight half of the code's designed distance. The decoding algorithm is formulated around the so-called key equation, and the Berlekamp-Massey algorithm itself is explained by the theory of linear recurring sequences. See any texts on coding theory \cite{macwilliams1983,huffman2003,blahut2003,lin2004,roth2006}.

Reed-Solomon codes are the simplest example of algebraic geometry codes---codes on the affine line, and generalized Reed-Solomon codes are a slight variation of Reed-Solomon codes. As alternant codes are defined as subfield subcodes of generalized Reed-Solomon codes, they inherit certain geometric structures from Reed-Solomon codes. From this point of view, the origin of the key equation and the linear recurring sequences in the theory of the decoding algorithm of alternant codes is somewhat mysterious.

Recently, in \cite{kwankyu2006}, it was shown that the decoding algorithm of Reed-Solomon codes using the Berlekamp-Massey algorithm can be understood as a special case of Sudan's list decoding of Reed-Solomon codes \cite{sudan1997,guruswami1999}. This result hints that we may formulate the classical decoding algorithm of alternant codes in terms of interpolation and division as in list decoding. The aim of this paper is to make this idea explicit.

Fitzpatrick \cite{fitzpatrick1995} was the first to show that the theory of linear recurring sequences can be removed in formulating the  decoding algorithm of alternant codes, using Gr\"obner bases of modules instead. Going one step further from his work, we will replace the key equation with an interpolation, which fits better with the geometric viewpoint on alternant codes.
 
In Section 2, we review the definitions of alternant codes and related codes. See \cite{roth2006} for more detailed treatment of alternant codes. In Section 3, we observe that decoding of alternant codes essentially reduces to that of Reed-Solomon codes, and present a decoding algorithm using the theory of Gr\"obner bases of modules. In fact we use very little of the Gr\"obner bases theory, and recommend \cite{cox2005} for an introduction to the subject. In Section 4, the case of BCH codes is briefly treated.

\section{Alternant codes}

Alternant codes are defined as subfield subcodes of generalized Reed-Solomon codes. So we let $\F\subset\E$ be an extension of finite fields, and first define generalized Reed-Solomon codes over $\E$.

Let $n$ be a positive integer, and $\E[x]_n=\set{f\in\E[x]\mid\deg(f)<n}$. We fix a set $\ga=\set{\ga_1,\ga_2,\dots,\ga_n}$ of $n$ distinct points of $\E$. For $\ga$, the evaluation map $\ev:\E[x]_n\to\E^n$ is defined by 
\[
	f\mapsto(f(\ga_1),f(\ga_2),\dots,f(\ga_n)).
\]
Clearly $\ev$ is an isomorphism of vector spaces over $\E$. The inverse map $\ev^{-1}$ is given by Lagrange interpolation as follows. Define
\[
%\begin{equation}\label{equcss}
	\tilde h_i=\prod_{j=1,j\neq i}^n(x-\ga_j),\text{ and } h_i=\tilde h_i(\ga_i)^{-1}\tilde h_i
%\end{equation}
\]
such that $h_i(\ga_j)=1$ if $j=i$, and $0$ otherwise. So $\set{h_1,h_2,\dots,h_n}$ forms a basis of $\E[x]_n$. For any vector $v=(v_1,v_2,\dots,v_n)\in\E^n$, we define
\[
	h_v=\ev^{-1}(v)=\sum_{i=1}^nv_ih_i\in\E[x]_n.
\]
For an integer $1\le k\le n$, the Reed-Solomon code $\RS(\ga,k)$ is defined as
\[
	\RS(\ga,k)=\set{\ev(f)\mid \deg f(x)<k,f(x)\in\E[x]}.
\]
It is well known that $\RS(\ga,k)$ is an $[n,k,n-k+1]$ linear code over $\E$. For a set $u=\set{u_1,u_2,\dots,u_n}$ of nonzero elements in $\E$, the distortion map $\tau_u$ on $\E^n$ is defined by
\[
	(v_1,v_2,\dots,v_n)\mapsto(u_1v_1,u_2v_2,\dots,u_nv_n).
\]
Obviously $\tau_u$ is a linear automorphism on $\E^n$ preserving Hamming weights. Later we will use the notation $v'=\tau_u^{-1}(v)$ for $v\in\E^n$. Now the generalized Reed-Solomon code $\GRS(\ga,u,k)$ is defined to be
\[
    \GRS(\ga,u,k)=\tau_u(\RS(\ga,k))=\set{\tau_u\circ\ev(f)\mid \deg f(x)<k,f(x)\in\E[x]}.
\]
As an isomorphic image of $\RS(\ga,k)$ by $\tau_u$, the generalized Reed-Solomon code $\GRS(\ga,u,k)$ is an $[n,k,n-k+1]$ linear code over $\E$. Note that the set of codewords $\set{\tau_u\circ\ev(x^a)\mid 0\le a\le k-1}$ forms a basis of $\GRS(\ga,u,k)$. The matrix whose rows are these $k$ codewords is called the canonical generator matrix of $\GRS(\ga,u,k)$. The family of generalized Reed-Solomon codes contains
their own duals.

\begin{prop}\label{propcbz}
The dual of $\GRS(\ga,u,k)$ is $\GRS(\ga,v,n-k)$, where $v=\set{v_i}$ with $v_i^{-1}=u_i\tilde{h}_i(\ga_i)$ for $1\le i\le n$.
\end{prop}

\begin{proof}
Let $0\le a\le k-1$ and $0\le b\le n-k-1$. As $\ev$ is an isomorphism,
\[
	x^{a+b}=\ev^{-1}((\ga_1^{a+b},\ga_2^{a+b},\dots,\ga_n^{a+b}))=\sum_{i=1}^n\ga_i^{a+b}h_i.
\]
Comparing the coefficients of $x^{n-1}$ on both sides, we see
\[
	0=\sum_{i=1}^n\tilde{h}_i(\ga_i)^{-1}\ga_i^{a+b}=\tau_u\circ\ev(x^a)\cdot\tau_v\circ\ev(x^b),
\]
where the dot denotes the inner product on $\E^n$. This completes the proof.
\end{proof}

Finally the alternant code $\ALT_\F(\ga,u,k)$ is defined by
\[
	\ALT_\F(\ga,u,k)=\GRS(\ga,u,k)\cap\F^n=\tau_u(\RS(\ga,k))\cap\F^n.
\]
So $\ALT_\F(\ga,u,k)$ is a linear code over $\F$ of length $n$ and dimension $\le k$, since a basis of $\ALT_\F(\ga,u,k)$ over $\F$ is linearly independent also over $\E$. Clearly its minimum distance is at least $n-k+1$, which is called the designed distance of the alternant code $\ALT_\F(\ga,u,k)$. 

\section{Decoding algorithm}

It is obvious that a decoding algorithm of $\GRS(\ga,u,k)$ correcting errors up to half of its minimum distance is immediately a decoding algorithm of $\ALT_\F(\ga,u,k)$ correcting errors up to half of its designed distance. In turn, a decoding algorithm of $\GRS(\ga,u,k)$ will be obtained by a slight modification of a decoding algorithm of $\RS(\ga,k)$. Below we present a decoding algorithm of $\GRS(\ga,u,k)$, and hence of $\ALT_\F(\ga,u,k)$, correcting up to $\fl{(n-k)/2}$ errors.

Let $c$ denote a codeword of $\ALT_\F(\ga,u,k)$ sent through a noisy channel. Suppose that $r=c+e$ is the received vector with error vector $e$. Let $t=\wt(e)$. We assume $2t<n-k+1$ so that $c$ is the unique codeword that lies in the Hamming sphere of radius $\fl{(n-k)/2}$ centered at $r$. Let $\E[x,y]_1=\set{f\in\E[x,y]\mid\ydeg(f)\le 1}$. Note that $\E[x,y]_1$ is a free module of rank $2$ over $\E[x]$. We consider 
\[
	M=\set{f\in\E[x,y]_1\mid\text{$f(\ga_i,u_i^{-1}r_i)=0$ for $1\le i\le n$}}.
\]
It is clear that $M$ is an $\E[x]$-submodule of $\E[x,y]_1$.

We review the necessary theory of Gr\"obner bases of submodules of $\E[x,y]_1$. Note that $x^iy^j$ with $i\ge 0$ and $j=0$ or $1$ are all the monomials of $\E[x,y]_1$. Given a parameter $s$, we define the monomial order $>_s$ as follows. The weights of the variables $x$ and $y$ are set to be $1$ and $s$, respectively, so that the $s$-weighted degree of the monomial $x^iy^j$ is $i+js$. Monomials are ordered by their weighted degree and if tied, the monomial with $y$ factor dominates the other. The minimal element of a submodule $S$ of $\E[x,y]_1$ with respect to $>_s$ is the element of $S$ with the smallest leading term, determined up to a constant. The following is trivial by Buchberger's $S$-pair criterion.

\begin{prop}
If $\set{f_1,f_2}$ is a basis of a submodule $S$ of $\E[x,y]_1$ with $\ydeg(f_1)=0$ and $\ydeg(f_2)=1$, then $\set{f_1,f_2}$ is a Gr\"obner basis of $S$.
\end{prop}

For the received vector $r$, define
\[
	h_{r'}=\sum_{i=1}^nr_iu_i^{-1}h_i,\quad \eta=\prod_{i=1}^n(x-\ga_i)
\]
Clearly $y-h_{r'},\eta\in M$. In fact,

\begin{prop}
$\set{\eta,y-h_{r'}}$ is a module basis of $M$.
\end{prop}

\begin{proof}
Let $ay+b\in M$ with $a,b\in\E[x]$. Note that $ay+b-a(y-h_r)=b+ah_r\in M\cap\E[x]$. Therefore $b+ah_r$ vanishes on $\ga_i$ for $1\le i\le n$, and we can write $b+ah_r=c\eta$ for some $c\in\E[x]$. Thus $ay+b=a(y-h_r)+c\eta$.
\end{proof}

Let $f_e=\prod_{e_i\not=0}(x-\ga_i)$. Suppose $c=\tau_u\circ\ev(h_{c'})$ with $\deg(h_{c'})<k$. Observe that $f_e(y-h_{c'})$ is in $M$. Moreover,
\begin{prop}
$f_e(y-h_{c'})$ is the minimal element of $M$ with respect to $>_{k-1}$.
\end{prop}

\begin{proof}
Assume $f_e(y-h_{c'})$ is not minimal in $M$. Then for some $a,b\in\E[x]$ not both zero,
\[
	\LT(f_e(y-h_{c'}))>_{k-1}\LT(a(y-h_{r'})+b\eta).
\]
Note that the $(k-1)$-weighted degree of $f_e(y-h_{c'})$ is $t+k-1$, and thus either 
\[
	t+k-1>\deg(a)+k-1\ge\deg(-ah_{r'}+b\eta)
\]
or
\[
	t+k-1\ge\deg(-ah_{r'}+b\eta)>\deg(a)+k-1.
\]
In either case, it follows that 
\[
	t>\deg(a),\quad t+k-1\ge\deg(-ah_{e'}+b\eta)
\]
since $h_{r'}=h_{c'}+h_{e'}$ and $k-1\ge\deg(h_{c'})$. We see that $a$ is nonzero by the second inequality. If 
$ah_{e'}=b\eta$, then $a(\ga_i)=0$ whenever $e_i\not=0$ so that $\deg(a)\ge t$, contradicting the first inequality. Hence $-ah_{e'}+b\eta$ is a nonzero polynomial in $x$. Note that it has at least $n-t$ zeros. Therefore 
\[
	t+k-1\ge\deg(-ah_{e'}+b\eta)\ge n-t.
\]
This contradicts our assumption that $2t<n-k+1$. 
\end{proof}

Observe that the minimal element $f_e(y-h_{c'})$ of $M$ with respect to $>_{k-1}$ should appear as an element of the Gr\"obner basis of $M$ with respect to $>_{k-1}$. Therefore once the Gr\"obner basis is at hand, the sent codeword $c$ can be retrieved by computing $\tau_u\circ\ev(h_{c'})$. Below we describe an algorithm converting the basis $\set{\eta,y-h_{r'}}$ to a Gr\"obner basis of the module $M$ with respect to $>_{k-1}$. 

Suppose that $A,B,C,D\in\E[x]$ such that
\[
	\set{Ay+B,Cy+D}
\]
is a basis of $M$. Assume that $\deg(B)+\deg(C)>\deg(A)+\deg(D)$ and that $\deg(A)+k-1<\deg(B)$, that is, $\ydeg(\LT(Ay+B))=0$. 

If $\deg(C)+k-1\ge\deg(D)$, that is, $\ydeg(\LT(Cy+D))=1$, then $\set{Ay+B,Cy+D}$ is a Gr\"obner basis of $M$. Suppose that $\deg(C)+k-1<\deg(D)$, and let $d=\deg(D)-\deg(B)$ and $c=\LC(D)\LC(B)^{-1}$. We now consider the following two cases. 

\begin{case1}
In this case,
\[
	\set{Ay+B,(C-cx^dA)y+(D-cx^dB)}
\]
is clearly a basis of $M$. Moreover 
\begin{itemize}
\item[(i)] $\deg(B)+\deg(C-cx^dA)>\deg(A)+\deg(D-cx^dB)$,
\item[(ii)] $\deg(D-cx^dB)-\deg(C-cx^dA)<\deg(D)-\deg(C)$.
\end{itemize}
\end{case1}

\begin{proof}
Note that by our assumption,
\[
	\deg(x^dA)=\deg(D)-\deg(B)+\deg(A)<\deg(C),
\]
and that $c,d$ were chosen such that
\[
	\deg(D-cx^dB)<\deg(D).
\]
We can easily check the assertions from these facts.
\end{proof}

\begin{case2}
In this case,
\[
	\set{Cy+D,(x^{-d}C-cA)y+(x^{-d}D-cB)}
\]
is a basis of $M$. Moreover
\begin{itemize}
\item[(iii)] $\deg(D)+\deg(x^{-d}C-cA)>\deg(C)+\deg(x^{-d}D-cB)$,
\item[(iv)] $\deg(x^{-d}D-cB)-\deg(x^{-d}C-cA)<\deg(D)-\deg(C)$.
\end{itemize}
\end{case2}

\begin{proof}
The assertions follow similarly from the facts that 
\[
	\deg(x^{-d}C)=\deg(B)-\deg(D)+\deg(C)>\deg(A),
\]
and that $\deg(x^{-d}D-cB)<\deg(B)$.
\end{proof}	

By (i) and (iii), we see that the above procedure can be iterated with the new basis given above in two cases, until 
$\deg(C)+k-1\ge\deg(D)$. The last condition eventually holds because (ii) and (iv) imply that the gap between the $(k-1)$-weighted degrees of $Cy$ and $D$ diminishes in each iteration. Hence we proved the following algorithm.

\begin{algD} Given the received vector $r=(r_1,r_2,\dots,r_n)$, this algorithm finds the sent codeword $c$ if there are at most $\fl{n-k}$ errors in $r$. The polynomials $\eta=\prod_{j=1}^n(x-\ga_j)$ and $u_i^{-1}h_i$ for $1\le i\le n$ are precomputed.

\begin{enumerate}
\item[D1.] Compute $-h_{r'}=-\sum_{i=1}^nr_iu_i^{-1}h_i$.
\item[D2.] Set
\[
	A\leftarrow 0,\quad B\leftarrow \eta,\quad
	C\leftarrow 1,\quad D\leftarrow -h_{r'}.
\]
\item[D3.] If $\deg(C)+k-1\ge\deg(D)$, then go to step D6.
\item[D4.] Set $d\leftarrow\deg(D)-\deg(B)$ and  $c\leftarrow\LC(D)\LC(B)^{-1}$.
\item[D5.] If $d\ge 0$, then set 
\[
	C\leftarrow C-cx^dA,\quad
	D\leftarrow D-cx^dB.
\]
If $d<0$, then set, storing $A$ and $B$ in temporary variables,
\[
\begin{aligned}
	A\leftarrow C,
		&\quad B\leftarrow D,\quad
	C\leftarrow x^{-d}C-cA,
		&\quad D\leftarrow x^{-d}D-cB.
\end{aligned}
\]
Go back to step D3.
\item[D6.] Output $\tau_u\circ\ev(-D/C)$ and the algorithm terminates.  
\end{enumerate}
\end{algD}

Alternatively we may use the Euclidean algorithm when we compute the new basis in the iteration steps, and obtain the algorithm below. We omit its proof, which can be found in \cite{kwankyu2006}.

\begin{algE} 
This algorithm performs the same task as Algorithm D, but depends on the Euclidean algorithm.

\begin{enumerate}
\item[E1.] Compute $-h_{r'}=-\sum_{i=1}^nr_iu_i^{-1}h_i$.
\item[E2.] Set
\[
	A\leftarrow 0,\quad B\leftarrow \eta,\quad
	C\leftarrow 1,\quad D\leftarrow -h_{r'}.
\]
\item[E3.] If $\deg(C)+k-1\ge\deg(D)$, then go to step E6.
\item[E4.] Compute $Q$ and $R$ such that $B=QD+R$, $\deg(R)<\deg(D)$ by the Euclidean algorithm.
\item[E5.] Set, storing $A$ in a temporary variable
\[
	A\leftarrow C,\quad B\leftarrow D,\quad
	C\leftarrow A-QC,\quad D\leftarrow R.
\]
Go back to step E3.
\item[E6.] Output $\tau_u\circ\ev(-D/C)$ and the algorithm terminates.  
\end{enumerate}
\end{algE}

\section{BCH codes}

Let $\gb$ be a primitive $n$th root of unity, which lies in an extension field
$\E$ of $\F$. For $b\in\Z$ and $1<\gd\le n$, the BCH code $\BCH(n,\gd,b)$ is
defined by
\[
    \BCH(n,\gd,b)=\set{f(x)\in\F[x]_n\mid\text{$f(\gb^i)=0$ for $b\le i<
    b+\gd-1$}},
\]
where we identify $f(x)=c_1+c_2x+\dots+c_nx^{n-1}$ with $(c_1,c_2,\dots,c_n)\in\F^n$.
Note that by definition, $(c_1,c_2,\dots,c_n)\in\F^n$ is a codeword of
$\BCH(n,\gd,b)$ if and only if
\[
    \begin{bmatrix}
        1   & \gb^b & \gb^{2b} & \cdots & \gb^{(n-1)b} \\
        1   & \gb^{(b+1)} & \gb^{2(b+1)} & \cdots & \gb^{(n-1)(b+1)} \\
        \vdots & \vdots & \vdots & \ddots & \vdots \\
        1   & \gb^{(b+\gd-2)} & \gb^{2(b+\gd-2)} & \cdots & \gb^{(n-1)(b+\gd-2)} \\
    \end{bmatrix}
    \begin{bmatrix}
    c_1 \\ c_2 \\ \vdots \\ c_n
    \end{bmatrix}=0.
\]
Observe that the matrix shown above is identical with the canonical generator
matrix of $\GRS(\ga,u,\gd-1)$ over $\E$ with
\[
    \ga=\set{1,\gb,\gb^2,\dots,\gb^{n-1}},\quad u=\set{1,\gb^b,\gb^{2b},\dots,\gb^{(n-1)b}}.
\]
This means that $\BCH(n,\gd,b)$ can be viewed as the subfield subcode over $\F$
of the dual code of $\GRS(\ga,u,\gd-1)$ over $\E$. By Proposition \ref{propcbz},
we see that
\[
	\BCH(n,\gd,b)=\GRS(\ga,v,n-\gd+1)\cap\F^n
\]
where $v=\set{n^{-1}\gb^{n-i(b+1)}\mid 0\le i\le n-1}$. Therefore this BCH code is an alternant code with designed distance $\gd$, which is also called the designed distance of the BCH code. As BCH codes are alternant codes, the decoding algorithm in the preceding section works for BCH codes.

\section{Conclusion}

As noted in the Introduction, the decoding algorithm in Section 3 is equivalent to the Berlekamp-Massey algorithm. In particular, the interpolation in D1 of Algorithm D corresponds to the syndrome computation. Therefore the decoding algorithm of alternant codes that we described in this paper is nothing but a disguise of the classical decoding algorithm based on the Berlekamp-Massey algorithm, or \textit{vice versa}. However, historically the classical decoding algorithm was first invented for BCH codes, and later found to work for general alternant codes. I believe that this historical accident has obscured the underlying principle of the decoding algorithm. Now Sudan's insight permits us to perceive that the classical decoding algorithm of alternant codes is in principle based on the properties of Reed-Solomon codes, but also works for BCH codes by accident.

\section*{Acknowledgements} I would like to thank Michael E.~O'Sullivan for his valuable comments and suggestions.

%\bibliographystyle{abbrv}
%\bibliography{IEEEabrv,../../bibliographies/CodingTheory}

\end{document}